\documentclass[12pt]{article}
\usepackage{amsfonts,amssymb}
\newtheorem{theorem}{Theorem}[section]

\newtheorem{definition}[theorem]{Definition}
\makeatletter
\@addtoreset{equation}{section}

\makeatother
\title{On a Poisson reduction for\\
Gel'fand--Zakharevich
manifolds\footnote{Ref. SISSA 31/2001/FM}}
\author{Gregorio Falqui\\
SISSA, Via Beirut 2/4, I-34014 Trieste, Italy\\
falqui@sissa.it\\[2ex]
Marco Pedroni\\
Dipartimento di Matematica, Universit\`a di Genova\\
Via Dodecaneso 35, I-16146 Genova, Italy\\
pedroni@dima.unige.it}

\begin{document}

\newtheorem{remark}{Remark}
\newenvironment{rem}{\begin{remark} \rm}{\end{remark}}

\maketitle
\begin{abstract}
We formulate and discuss a reduction theorem for Poisson pencils associated
with a class of integrable  systems, defined on bi-Hamiltonian manifolds,
recently studied by Gel'fand and
Zakharevich. The reduction procedure is suggested by
the bi-Hamiltonian approach to the  Separation of Variables problem.
\end{abstract}

\noindent
{\bf Key words:} Bi-Hamiltonian manifolds, Hamiltonian reduction, Poisson
manifolds.

\section{Introduction}
The aim of this paper is to present and prove some results about Poisson
reduction for bi-Hamiltonian manifolds. The methods presented in the paper
are an outgrowth of
a geometric theory of Separation of Variables, based on the notion of
bi-Hamiltonian
geometry, introduced in recent years (see, e.g., \cite{mt97,Bl98,FMT00}),
which is thoroughly discussed in~\cite{FP1} and~\cite{FMP2}.
The cornerstones of such a theory are the (related) concepts of $\omega N$
manifold and
of bi-Lagrangian foliation. An $\omega N$ manifold $M$ can be viewed as a
special  bi-Hamiltonian manifold,
where one of the two
compatible Poisson brackets defined on $M$ is actually {\em symplectic},
i.e., is associated with a symplectic 2-form $\omega$. A
bi-Lagrangian foliation $\cal F$ defined on an $\omega N$ manifold $M$ is a
foliation of $M$ which is {\em
  Lagrangian} with respect to both brackets. The content of the
``bi-Hamiltonian
theorem of Separation of Variables'' can be (roughly) summarized as follows:
\\
A Hamiltonian $H$ is {\em separable} in a set of coordinates (called
Darboux-Nijenhuis coordinates)
naturally associated with the $\omega N$ structure of $M$ if and only if the
Hamiltonian vector field $X_H$, defined by the equation
\[
\omega(X_H,\cdot)=-dH\ ,
\]
is tangent to a bi-Lagrangian foliation.

Although the notions of Hamiltonian vector field (and, {\em a fortiori}, of
separability)
pertain to the category of {\em symplectic manifolds}, it
``experimentally'' turns out that a major source of separable systems admit
a more
natural formulation in a wider class of manifolds, that is, Poisson
manifolds.
This feature is particularly evident in the case of integrable systems
connected with soliton equations and loop algebras
(see, e.g., \cite{FMcL,DKN,Ha1,RSTS,Sk95}), related to the theory of Lax
equations and the $R$-matrix formalism.

We will be interested in the class of systems nowadays
known as {\em Gel'fand-Zakharevich} (GZ) systems.
These are bi-Hamiltonian systems defined over a
bi-Hamiltonian manifold
$M$ where none of the Poisson brackets is symplectic,
over which, so to speak,  the geometry of
the Poisson pencil $P'-\lambda P$ itself selects a complete family
of mutually commuting Hamiltonians.
Indeed, by definition, for a (torsionless) GZ system of rank
$k$, this family can be
grouped in $k$ {\em Casimirs  of the Poisson pencil\/}.
A Casimir of the pencil is a polynomial
$H(\lambda)=H_0\lambda^n+H_1\lambda^{n-1}+\cdots+ H_n$,
whose coefficients are functions on
$M$, that satisfy the equation
\begin{equation}
  \label{eq:defcasimir}
  (P'-\lambda P) d H(\lambda)=0\>.
\end{equation}
It is well-known (and quite easy to check)
that such equation entails the Lenard
recursion relations
for the vector fields $X_i=P d H_{i+1}$, and, as a consequence, the mutual
commutativity with respect to both brackets of the coefficients
$H_i$. So, the family of the vector fields associated with the $k$
above-mentioned Casimirs defines a distribution $\cal X$ in $TM$
(called the axis of the pencil) which has the following property: it is
tangent to the symplectic leaves $\bar S$ of any fixed element $\bar P$ of
the Poisson
pencil defined on $M$, and, when viewed as a distribution defined on the
symplectic manifold $\bar S$ (endowed with the natural symplectic structure
induced by $\bar P$) it is a Lagrangian distribution. Once we have fixed
such
a ``preferred'' element $\bar P$, we can thus discuss whether we can induce
on
$\bar S$ another Poisson bracket,
starting from another element of the Poisson
pencil $P_\lambda$. In doing this, we require (for the reasons briefly
addressed
above, and related to the Separation of Variables problem) to preserve two
properties:

a) compatibility of the reduced brackets on $\bar S$;

b) commutativity of the (restriction to $\bar S$ of the) Hamiltonians $H_j$.

To this end, we will have to make an assumption on the pencil, that is, we
will
suppose that $P_\lambda$ admits a distribution $\cal Z$ transversal to the
symplectic
leaves of $\bar P$, enjoying ``good'' properties (to be discussed later on
in the paper).

The problem can be tackled in two ways: from the point of view of
Poisson tensors (on a tubular neighbourhood of $\bar S$)
and from the point of view of induced 2-forms (on the single leaf $\bar S$).
Since, in
our opinion, both ways have their own strong points we decided to divide the
paper in two sections, and present both points of view. In particular, the
``Poisson'' point of view allows for simpler proofs and a nice
description in terms of Poisson-Lichnerowicz geometry. The other point of
view has the advantage that it clearly points out that the problem ``lives''
on single symplectic leaves, and frames its study within the scheme of
symplectic geometry.

The results contained in this paper have been used (more specifically those
concerning what we call the {\em strong form of the reduction theorem} in
Remark 1) in \cite{FMT00,FMPToda,DM01}.
\par\smallskip\noindent
{\bf Acknowledgments.} The results presented in this paper have been
obtained in
collaboration with Franco Magri, to whom we are very
grateful. We wish to thank
also the organizers for the nice atmosphere
at the conference. This work was partially supported by INdAM--GNFM and the
Italian M.I.U.R. under the research project {\em Geometry of
Integrable Systems}.

\section{GZ Poisson pencils and their reduction to symplectic leaves}
\label{sec:gz}
We are interested in a special class of  GZ systems, known
as complete torsionless rank $k$ systems of pure Kronecker type.
They are studied in, e.g., \cite{GZ93,GZ99,Pa99,Tu99,Za99}).

They can be  defined as the datum of:\\
i) a bi-Hamiltonian manifold $(M,\{\cdot,\cdot\}_\lambda)$, that is, a
manifold $M$ endowed with a
  linear pencil of Poisson brackets $\{\cdot,\cdot\}_\lambda=
\{\cdot,\cdot\}'-\lambda\{\cdot,\cdot\}$ or equivalently with a linear
pencil $P_\lambda=P'-\lambda P$ of Poisson tensors, where
\[
\{f,g\}_\lambda=\langle df, P_\lambda dg\rangle.
\]
ii) a collection of $k$ polynomial Casimir functions
$ H^{(1)},\ldots, H^{(k)},$
that is, a collection of degree $n_j$ polynomials
\[
H^{(j)}(\lambda)=\sum_{i=0}^{n_j} H^{(j)}_{i}\lambda^{n_j-i}
\]
such that:\par
a) $n_1+n_2+\cdots+n_k=n$, with $\mbox{dim}\,M=2n+k$.\par
b) The differentials $\{d H^{(j)}_s\}_{j=1,\ldots k;\>
s=0,\ldots, n_j}$ are linearly independent at every point and so define an
$(n+k)$--dimensional distribution in $T^*M$. \par
The collection of the $n$ bi-Hamiltonian vector fields
\begin{equation}
  \label{eq:01}
  X^{(j)}_k=P d\,H^{(j)}_{k+1}=P' d\,H^{(j)}_k
\end{equation}
associated with the Lenard sequences defined by the  polynomials Casimir
$H^{(j)}$ is called the {\em GZ system} associated with the given GZ
manifold,
or {\em axis} of the bi-Hamiltonian manifold\ $M$.
We now consider (clearly, this specific choice is inessential) the preferred
Poisson tensor $\bar P$, mentioned in the Introduction, to be exactly $P$.
Let $S$ be one of its symplectic leaves.
We seek for a
deformation of the  Poisson structure $P'$ to $Q=P'+\Delta P'$,
such that the following three properties hold:
\begin{enumerate}
\item $Q$ restricts to $ S$;
\item $Q-\lambda P$ is still a Poisson pencil;
\item $P'-\lambda P$ and $Q-\lambda P$ share the same axis.
\end{enumerate}
To this end we consider a distribution $\cal  Z$ transversal to $ S$ and
such
that it splits the tangent space to $M$ as
\begin{equation}\label{eq:spl}
TM=T S\oplus\cal Z\ .
\end{equation}
We assume, at this stage, that $\cal Z$ is defined in a whole tubular
neighbourhood $U_ S$ of $ S$.
We consider the family of Casimir functions
$(H^{(1)}_0,\dots,H^{(k)}_0)$
of $\{\cdot,\cdot\}$ defining
$ S$, and a family of vector fields $(Z_1,\dots,Z_k)$ spanning $\cal  Z$. We
can assume that the transversal vector fields $Z_a$ are normalized:
\begin{equation}
\label{eq:nrmZ}
Z_a(H_0^{(b)})=\delta_a^b\> .
\end{equation}
We consider the ``first'' vector fields of the each Lenard sequence, that
is,
\[
X'_a=P'd\,H^{(a)}_0\>,
\]
and we define the new bivector
\begin{equation}
  \label{eq:pprimo}
  Q=P'-\sum_{a=1}^k X'_a\wedge Z_a\>.
\end{equation}
\begin{theorem}
\par
1) The bivector $Q$ restricts to $ S$, and all the GZ Hamiltonians
$H^{(a)}_l$
are skew orthogonal with respect to the second ``bracket'' defined by $Q$,
that is,
\begin{equation}\label{eq:fgq}
\{H^{(a)}_l,H^{(b)}_k\}_Q :=<dH^{(a)}_l,Q\, dH^{(b)}_k >=0\>.
\end{equation}
2) The second bracket satisfies the Jacobi identity if and only if
\begin{equation}\label{eq:scQQ}
\sum_{a=1}^k X'_a\wedge \left(L_{Z_a}(P')-\sum_{b=1}^k[Z_a,X'_b]\wedge
Z_b\right)-\frac12 \sum_{a,b=1}^k X'_a\wedge X'_b\wedge [Z_a,Z_b]=0,
\end{equation}
that is, if and only if
\begin{equation}\label{eq:scQQ-2}
\sum_{a=1}^k X'_a\wedge \left(L_{Z_a}(Q)+\frac12
\sum_{b=1}^k X'_b\wedge [Z_a,Z_b]\right)=0.
\end{equation}
3) In this case, the two brackets define a
Poisson pencil if and only if
\begin{equation}\label{eq:scPQ}
\sum_{a=1}^k X'_a\wedge L_{Z_a}(P)=0\>.
\end{equation}
\end{theorem}
{\bf Proof}. To prove that $Q$ restricts to $ S$  we remark that, if we
consider
bivectors as maps from $T^*M$ to $TM$, the corresponding expression for the
map associated with $Q$ is given by
\[
Q(\alpha)=P'(\alpha)+\sum_{a=1}^k \left(<\alpha,X'_a> Z_a -<\alpha,Z_a>
  X'_a\right)\>.
\]
We must show that $\mbox{Im}(Q)\subset TS$, i.e., that for every
1-form $\alpha$ and $b=1,\ldots,k$,
\[
<dH^{(b)}_0, Q (\alpha)>=0\ .
\]
This is true thanks to the antisymmetry of $Q$, and the fact that
$QdH^{(b)}_0=0$. The validity of Eq.~(\ref{eq:fgq})
is proved exactly in the same way,
taking into account the commutativity of the vector fields entering the
Lenard sequences.

The proof of the last two assertions is done via a computation which makes
use
of the formal properties of the Schouten brackets of multivectors, and,
especially, of the fact that the Schouten bracket is an extension to
polyvector fields of the Lie derivative for vector fields (see,
e.g.~\cite{Vais}). In particular, we will use the following facts:\\
i) If $X$ and $Y$ are vector fields, the Schouten bracket $[X,Y]_S$
coincides
with the commutator $[X,Y]$. \\
ii) the Lie derivative along a vector field $Z$ of the
wedge product of two vector fields $X$ and $Y$
satisfies:
\[
L_Z(X\wedge Y)=[Z,X]\wedge Y+X\wedge[Z,Y]\>.
\]
iii) if $X$ is a vector field and $P$ a bivector, then
\[
[X,P]_S=L_X(P)\>.
\]
iv) If $X,Y$ are vector fields and $P$ is a bivector one has:
\[
[X\wedge Y, P]_S=Y\wedge[X,P]_S-X\wedge[Y,P]_S=
Y\wedge L_X(P)-X\wedge L_Y(P)\>.
\]
Using i) through iv) one can argue as follows:
It is well-known that $\{\cdot,\cdot\}_Q$ is a
Poisson bracket\ if and only if the Schouten bracket $\left[Q,Q\right]_S$
vanishes,
and that the
compatibility between $\{\cdot,\cdot\}_Q$ and
$\{\cdot,\cdot\}$ takes the form
$\left[Q,P\right]_S=0$.
Let us compute
\[
\begin{array}{rcl}
\left[Q,Q\right]_S &=& \left[P'-\sum_{a=1}^k X'_a\wedge Z_a,
P'-\sum_{a=1}^k X'_a\wedge Z_a\right]_S\\
&=& -2 \sum_{a=1}^k \left[X'_a\wedge Z_a,P'\right]_S+
\sum_{a,b=1}^k \left[X'_a\wedge Z_a,X'_b\wedge Z_b\right]_S\\
&=& -2 \sum_{a=1}^k \left(L_{X'_a}P'\wedge Z_a-X'_a\wedge
L_{Z_a}P'\right)\\ && +
\sum_{a,b=1}^k \left(2 X'_a\wedge [Z_a,X'_b]\wedge Z_b-X'_a\wedge
X'_b\wedge [Z_a,Z_b]+[X'_a,X'_b]\wedge Z_a\wedge Z_b\right)\ .
\end{array}
\]
But $L_{X'_a}P'=0$ and $[X'_a,X'_b]=0$,
so that
\[
\left[Q,Q\right]_S
= 2 \sum_{a=1}^k X'_a\wedge\left(L_{Z_a}P'-\sum_{b=1}^k
[Z_a,X'_b]\wedge
Z_b\right)-\sum_{a,b=1}^k X'_a\wedge
X'_b\wedge [Z_a,Z_b]\ .
\]
Therefore $\left[Q,Q\right]_S=0$ if and only if
(\ref{eq:scQQ}) (as well as its equivalent form (\ref{eq:scQQ-2})) holds.

As far as assertion 3) is concerned, we have
\[
\begin{array}{rcl}
\left[Q,{P}\right]_S &= &\left[P'-\sum_{a=1}^k X'_a\wedge
Z_a,P\right]_S= -\sum_{a=1}^k \left[X'_a\wedge Z_a,P\right]_S\\
&=& \sum_{a=1}^k \left(-L_{X'_a}P\wedge Z_a+X'_a\wedge L_{Z_a}P\right)
\ .
\end{array}
\]
Now we recall that the compatibility condition between $P$ and
$P'$ can be written as
\[
L_{P'dF} P+L_{PdF} P'=0\qquad \mbox{for all $F\in C^\infty(M)$.}
\]
Since $H_0^{(a)}$ is a Casimir of $P$, this implies that
$L_{X'_a}P=0$. Hence we have
\[
\left[Q,{P}\right]_S=\sum_{a=1}^k X'_a\wedge L_{Z_a}P
\ ,
\]
and the theorem is proved.\ \rule{5pt}{5pt}
\begin{rem}
When the transversal distribution $\cal  Z$ is integrable,
Equation~(\ref{eq:scQQ}) simplifies to
\begin{equation}\label{eq:sc2}
\sum_{a=1}^k X'_a\wedge \left(L_{Z_a}(P')-\sum_{b=1}^k[Z_a,X'_b]\wedge
Z_b\right)=0\>.
\end{equation}
So, the conditions for $Q-\lambda P$ to be a Poisson pencil reduce to
\begin{equation}
  \label{eq:sc000}
  \begin{array}{l}
\sum_{a=1}^k X'_a\wedge L_{Z_a}(Q)=0\\
\sum_{a=1}^k X'_a\wedge L_{Z_a}(P)=0\>.
\end{array}
\end{equation}
The ``strong'' solutions to this system, that is, the distributions $\cal
Z$
spanned by vector field\ satisfying, for $a=1,\ldots, k$,
\begin{equation}
  \label{eq:oldcond}
   \begin{array}{l}
L_{Z_a}P'=\sum_{b=1}^k [Z_a,X'_b]\wedge Z_b\\
L_{Z_a}P=0\>,
\end{array}
\end{equation}
can be described in the framework
of the Marsden--Ratiu reduction scheme~\cite{MR86}
for Poisson manifolds. Indeed,  one can notice  that the
conditions~(\ref{eq:oldcond}) on the Poisson pencil imply (actually, are
equivalent to, see \cite{Vais})  the fact that the ring of functions which
are left
invariant by the distribution $\cal  Z$ are  a {\em Poisson\/} subalgebra
with respect to
the pencil. This means that if $Z(F)=Z(G)=0$ for every $Z\in\cal  Z$, then
$Z(\{F,G\}_\lambda)=0$. Thus the bi-Hamiltonian structure can be projected
onto every symplectic leaf $S'\subset U_ S$ of $P$, and the restriction of
$Q$ to $ S'$ coincides with the
{\em reduction} of $P'$ to $ S'$. The reader should however be aware of the
fact that, in
the general case, that is, when $Q$ satisfies
Equations~(\ref{eq:scQQ-2}), the
reduction scheme herewith presented does not fit in the MR setting.
\end{rem}
\begin{rem} It is interesting to compare our approach to the reduction of
$P'$ with the classical
Dirac reduction procedure for second class constraints. The latter (see,
e.g.,\cite{MR86,Vais}) is
usually described as follows.  Let $(M,\{\cdot,\cdot\})$ be a
Poisson (or even
symplectic) manifold, and let $\{\phi_1,\ldots,\phi_{2k}\}$ be a family of
``constraints'' for a Hamiltonian system defined on $M$. One says that the
constraints are second class if the matrix of Poisson brackets
\begin{equation}
  \label{eq:D1}
  C_{ab}=\{\phi_a,\phi_b\}
\end{equation}
is nondegenerate on a submanifold $S\subset M$, where $S$ is defined by the
$2n$ equations $\phi_a=\mbox{const}_a, a=1,\ldots,2k$.
The Dirac bracket $\{\cdot,\cdot\}^D$ is defined on $S$ as follows:
\begin{equation}
  \label{eq:D2}
  \{F,G\}^D= \{F,G\}-\sum_{a,b=1}^n \{F,\phi_a\}(C^{-1})_{ab}\{\phi_b,G\}\>.
\end{equation}
 In terms of Poisson tensors it is not difficult to check that,
if $P$ is
 associated with $\{\cdot,\cdot\}$ and $X_a$ is the Hamiltonian vector field
associated
 with $\phi_a$, that is, $X_a=P d\phi_a$, the Poisson tensor associated with
$\{\cdot,\cdot\}^D$ is
 \begin{equation}
   \label{eq:D3}
   P^D:=P-\frac12 \sum_{a,b=1}X_b\wedge \big(C^{-1}\big)_{ab}X_a\>.
 \end{equation}
This can be interpreted (see also \cite{Vais}) as follows.  The Dirac
bracket is a deformation of the ``original one'' to one for which the
constraint functions (that define the special submanifold $S$, or, better, a
local foliation given by the constraints $\{\phi_a\}$) are Casimirs. In
particular,  the analogy with Eq.~(\ref{eq:pprimo}) is enhanced by remarking
that the vector fields
\[
Y_a:=\sum_{b=1}^k \big(C^{-1}\big)_{ab} X_a
\]
of Eq.~(\ref{eq:D3}) are normalized with respect to the Casimirs of $P^D$.
The fundamental difference between the two instances resides in the fact
that,
in our case,  the Poisson brackets of the functions $C_a$ is maximally
degenerated, that is, it is the {\em zero}
matrix. So, the choice of the transversal
distribution $\cal  Z$  is  a ``free input'' in our problem.
As a consequence of such a freedom, however, we are no longer
guaranteed that the reduced ``brackets'' are Poisson brackets, and so one
has
to impose on $\cal  Z$ the condition~(\ref{eq:scQQ}).
\end{rem}
\begin{rem}
The deformation $P'\to Q$ defined by
Eq.~(\ref{eq:pprimo}) is
not the unique satisfying the requirements that $Q$ restricts
to $S$ and
that the GZ Hamiltonians are in involution. For instance, one could consider
\[
Q'=Q+\Delta
\]
with $\Delta$ a section of the second exterior product of the axis $\cal X$
of
the GZ manifold. Correspondingly, the requirements that $Q'-\lambda P$ be a
Poisson
pencil would take a more complicated form. The choice we made can be
considered as a ``minimal'' one.
\end{rem}
\begin{rem}
\label{rem4}
One can notice the following: if
Eqs.~(\ref{eq:scQQ})
and~(\ref{eq:scPQ}) do not hold in the whole tubular neighbourhood $U_ S$
but,
say, on a single symplectic leaf\ $\bar S$, we can still say that such a
single leaf is
endowed with a bi-Hamiltonian structure, of regular type, since, by
definition,
$P\vert_{\bar S}$ is clearly symplectic. Actually, as we shall see in the
next section, one can also drop the assumption that the distribution $\cal
Z$ be
defined in $U_S$ and require it to exist on a single symplectic leaf. To do
that, it is convenient to tackle
the problem from a different point of view.
\end{rem}

\section{The $\mathbf{\omega N}$ point of view}
We want now to discuss the problem at hand
from the
point of view of symplectic geometry, or, to be more precise, from the point
of view of the geometry of $\omega N$ manifolds. As we have anticipated in
Remark \ref{rem4} above,
the advantage of this point of view is
that it makes clear that the assumptions about the existence of the
transversal distribution in a whole tubular neighbourhood of the symplectic
leaf $ S$ are not necessary, and that the reduction process can be
discussed,
so to say, ``leaf by leaf'', even if it involves more complicated
computations (which we spare to the reader).
To proceed, we need to recall in more details some notions.
The basic elements of the concept of $\omega N$ manifold
are a symplectic manifold $(M,\omega)$, of dimension $2n$, and a second
closed 2-form $\omega'$. To this form we associate the recursion operator
$N$ defined as
\begin{equation}
\label{3.2}
\omega'(X,Y)=\omega(NX,Y)\ .
\end{equation}
\begin{definition}{Definition}
We say that $M$ is an $\omega N$ manifold if the Nijenhuis torsion of $N$ is
vanishing, i.e.,
\begin{equation}
T_N(X,Y):=[NX,NY]-N[NX,Y]-N[X,NY]+N^2[X,Y]=0
\end{equation}
for all pairs $(X,Y)$ of vector fields on $M$. In this case we say, for
short, that
$N$ is the recursion operator of the manifold $M$.
\end{definition}

An alternative form of the vanishing condition of the torsion of $N$ is
obtained by introducing the third 2-form $\omega''$ defined by
\begin{equation}
\omega''(X,Y)=\omega(NX,NY)\ .
\end{equation}
Indeed, it turns out that the torsion of $N$ vanishes if and only if
$\omega''$ is closed. To check this, it suffices to use the identity
\begin{equation}
\label{3.5b}\begin{array}{rcl}
d\omega''(X,Y,Z)&=&d\omega'(NX,Y,Z)+ d\omega'(X,NY,Z)-d\omega(NX,NY,Z)\\
&&-\omega(T_N(X,Y),Z)\ ,
\end{array}
\end{equation}
relating the exterior derivatives of the 2-forms $\omega$, $\omega'$,
$\omega''$ and
the torsion $T_N$ of the recursion operator.
The most direct link between the theory of $\omega N$ manifolds and
bi-Hamiltonian manifolds is the
following: If $Q-\lambda P$ is a Poisson pencil on a bi-Hamiltonian manifold
$M$, and if $P$ is
symplectic, with symplectic form $\omega$, the $(1,1)$ tensor field
\[
N=Q\cdot P^{-1}
\]
has vanishing Nijenhuis\ torsion, so that the pair $(\omega,N)$ endows $M$
with the structure of an $\omega N$ manifold. We refer to \cite{K-SM,MM} for
fuller details
and proofs.

Let us consider a Poisson manifold $M$. It is well-known that such a
manifold can be
seen as a ``glueing'' of symplectic leaves.
Let $P$ the Poisson bivector corresponding to the Poisson bracket, that is,
$\{F,G\}=:\langle dF,P dG\rangle$, and let us denote with
\begin{equation}
\label{3.7}
X_F=P dF
\end{equation}
the Hamiltonian vector field associated with the function $F$.
As it is well known, the leaves
of the distribution generated by the vector fields
$X_F, F\in C^\infty(M)$,
are endowed with the a canonical symplectic form $\omega$,
explicitly defined as
\begin{equation}
\label{symfor}
\omega(X_F,X_G):=\{F,G\}\ .
\end{equation}
Let us suppose now that $M$ be endowed with a second Poisson bracket,
denoted with $\{\cdot,\cdot\}'$, which we assume to be compatible with the
Poisson bracket associated with $P$.
In the $\omega N$ setting, it is thus natural to discuss
how to use the second Poisson bracket to induce a second
closed 2-form $\omega'$ on a fixed symplectic leaf $S$ of the first bracket,
in such a
way to provide $S$ with an $\omega N$ manifold structure. Since the
symplectic leaves of
$\{\cdot,\cdot\}'$ are not contained, in general, in the ones of the first
bracket,
we cannot use the analog of (\ref{symfor}) in order to define $\omega'$.

To
attain our aim, we must be able to ``project'' in a suitable way the
Hamiltonian
vector fields $X'_F$ (that is, the Hamiltonian vector fields associated with
$\{\cdot,\cdot\}'$)
on the symplectic leaf $S$. As the
projection process has a local character, without loss of generality we
can restrict to an open subset where the functions
$(C_1,\dots,C_k)$ form a basis in the algebra of
Casimir functions of $\{\cdot,\cdot\}$. In the intersection of such an open
set with $S$ we consider a family of $k$
vector fields $Z_a$, $a=1,\dots,k$, which, at the points $p\in  S$,
span a subspace of $T_pM$ complementary (and hence
transversal) to $T_pS$. We still agree to normalize these vector fields as
\begin{equation}
\label{3.10}
Z_a(C_b)=\delta_{ab}
\end{equation}
on $ S$.
As in Section~\ref{sec:gz},
we use the vector fields $Z_a$ to split, in every point $p$ of $ S$, the
tangent
space to $M$ as the direct sum
\begin{equation}
T_pM=T_p S\oplus {\cal  Z}_p
\end{equation}
of the  tangent space to the leaf and of the distribution spanned by the
vector fields $Z_a$. Dually, we split the cotangent space as the direct sum
\begin{equation}
\label{split}
T^*_pM={\cal  Z}_p^0 \oplus (T_p S)^0
\end{equation}
of the annihilators. We identify the annihilator ${\cal  Z}^0$ of $\cal  Z$
with the
cotangent space of the symplectic leaf, and we denote with $\pi^*$ the
canonical
projection on ${\cal  Z}^0$ in the splitting (\ref{split}). One has that
\begin{equation}
\pi^*(dF)=dF-\sum_{a=1}^k Z_a(F) dC_a\ .
\end{equation}
We still denote with $P'$ the bivector associated with the second Poisson
bracket, and we
introduce a second 2-form $\omega'$ on $ S$ according to the relation
\begin{equation}
\omega'(X_F,X_G):=\langle P'\circ\pi^*(dF),\pi^*(dG)\rangle\> .
\end{equation}

The problem we have to discuss now is how to choose the vector fields $Z_a$
in
such a way that the 2-form $\omega'$ define an $\omega N$ structure on the
selected leaf $ S$. We still work under the assumption that
Casimirs $C_a$ of the first bracket form an Abelian algebra with respect to
the second bracket, i.e., that $\{C_a,C_b\}'=0$. It implies
that the  vector fields $X'_a=P' dC_a$ are tangent to
$ S$ and, therefore, it is not necessary to project these vector fields onto
$ S$
(whence the simplification of some of the formulae in the sequel). In
particular, for $\omega'$ we obtain  the expression
\begin{equation}
\label{3.16}
\omega'(X_F,X_G)=\{F,G\}'+\sum_{a=1}^k \left(X'_a(F)Z_a(G)-
X'_a(G)Z_a(F)\right)\>,
\end{equation}
to be compared with formula~(\ref{eq:pprimo}).

According to the definitions given at the beginning of this section,
we have to insure that the 2-form $\omega'$ is
closed.

\begin{theorem}
\label{pfr1}
The 2-form $\omega'$ is closed if and only if
\begin{equation}
\label{3.17}
\sum_{a=1}^k X'_a\wedge L_{\tilde Z_a}(P)=0
\end{equation}
at the points of $ S$, where $\tilde Z_a$ is {\em any}
extension of the vector field $Z_a$ (which are in principle defined only on
$ S$)
to a tubular neighbourhood of $ S$ in $M$.
\end{theorem}
{\bf Proof.} We have to compute the exterior derivative of $\omega'$ on $
S$.
Thanks to the Palais formulae, one has
\begin{equation}
\label{3.18}
d\omega'(X_F,X_G,X_H)=\sum_{\mbox{cyclic}}\left(X_F
\omega'(X_G,X_H)-\omega'([X_F,X_G],
X_H)\right)\ .
\end{equation}
Before proceeding to the computation of this exterior derivative, we recall
that the compatibility condition between the two Poisson brackets can be
written as
\begin{equation}
\label{3.19}
\sum_{\mbox{cyclic}}\left(\{F,\{G,H\}\}'+\{F,\{G,H\}'\}\right)=0
\end{equation}
for every triple of functions $(F,G,H)$ on $M$. A particular case of this
identity is
\begin{equation}
\label{3.20}
\{F,\{G,C_a\}'\}+\{G,\{C_a,F\}'\}+\{C_a,\{F,G\}'\}=0\ ,
\end{equation}
obtained by putting $H=C_a$. Now let us substitute the expression of
$\omega'$
in Equation (\ref{3.18}). Let us use the definition (\ref{3.7}) of the
vector field\
$X_F$ and the commutation property $[X_F,X_G]=X_{\{F,G\}}$ of the
Hamiltonian\
vector fields.
Collecting in a suitable way the different terms, using the identities
(\ref{3.19}) and (\ref{3.20}),
we get straight away the identity
\begin{equation}
d\omega'(X_F,X_G,X_H)= \sum_{a=1}^k \left(X'_a\wedge L_{\tilde
Z_a}(P)\right) (dF,dG,dH)\ .
\end{equation}
It shows that the exterior derivative of the 2-form $\omega'$, evaluated on
the
Hamiltonian\ vector fields $X_F$, $X_G$ ,$X_H$, coincides with the 3-vector
$\sum_{a=1}^k X'_a\wedge L_{\tilde{Z}_a}(P)$ evaluated on the
differentials of the
corresponding Hamiltonians. This completes the proof.\
\rule{5pt}{5pt}

The second step is to check under which additional hypotheses the torsion of
the recursion operator $N$, associated with $\omega'$, vanishes too. To this
aim, we will write explicitly $N$. From (\ref{3.16}), taking into account
that $\{C_a,C_b\}'=0$, one finds that
\begin{equation}
\label{3.21}
NX_F=X'_F+\sum_{a=1}^k \left(X'_a(F) Z_a-Z_a(F) X'_a\right)\ .
\end{equation}
\begin{theorem}
Suppose that $d\omega'=0$ on $ S$.
Then the torsion of $N$ vanishes if and only if, at the points of S,
\begin{equation}
\sum_{a=1}^k X'_a\wedge R_a=0,
\end{equation}
where
\begin{equation}\label{eq:neweq}
R_a=L_{\tilde Z_a}(P'-\sum_{b=1}^k X'_b\wedge \tilde
Z_b)+\frac12\sum_{b=1}^k
X'_b\wedge [\tilde{Z}_a,\tilde{Z}_b],
\end{equation}
and, as above, $\tilde Z_a$ is {\em any}
extension of  $Z_a$ to a tubular neighbourhood of $ S$ in $M$.
\end{theorem}
{\bf Proof.} The proof of this assertion is, in ths formalism,
rather long, even if in principle not
difficult. The reason is that the expression of the torsion of
$N$ contains several terms that must be assembled and managed carefully.
Thus, we will give the essential lines of the computation, omitting the
details. We begin with computing the expression
\begin{equation}
[X_F,X_G]_N:=[NX_F,X_G]+[X_F,NX_G]-N[X_F,X_G]\ .
\end{equation}
Then we plug in it the form (\ref{3.21}) of $NX_F$, we compute the Lie
brackets and we make use once again of (\ref{3.19}) and (\ref{3.20}) to
simplify the resulting expression. Using also the condition
 (\ref{3.17}) of
closedness of the 2-form $\omega'$, which we already assumed,
we arrive in this way at the expression
\begin{equation}
\begin{array}{rcl}
[X_F,X_G]_N :&=& X_{\{F,G\}'}
+\sum_{a=1}^k \left(Z_a(G) X_{X'_a(F)}- Z_a(F) X_{X'_a(G)}\right)\\ &&\quad
+\sum_{a=1}^k \left(X'_a(F) X_{Z_a(G)}- X'_a(G) X_{Z_a(F)}\right)\
.\end{array}
\end{equation}
It shows that $[X_F,X_G]_N$ is a linear combination of Hamiltonian (with
respect to $\{\cdot,\cdot\}$) vector fields. This make simpler the
calculation of $N[X_F,X_G]_N$.
Finally, we evaluate the Lie bracket $[NX_F,NX_G]$. Subtracting
$N[X_F,X_G]_N$ to this Lie bracket, assembling with care the large number of
terms of the previous expressions, using the identity
\begin{equation}
[X'_F,X'_G]=X'_{\{F,G\}'}
\end{equation}
and the involutivity of the Casimir functions $C_a$,
we obtain a still quite complicated expression for the torsion
\begin{equation}
T_N(X_F,X_G)= [NX_F,NX_G]-N[X_F,X_G]_N
\end{equation}
we want to compute. Finally, evaluating the symplectic scalar product
\begin{equation}
\omega(X_H,T_N(X_F,X_G))=\langle T_N(X_F,X_G),dH\rangle\ ,
\end{equation}
we get the final identity
\begin{equation}
\omega(X_H,T_N(X_F,X_G))= \sum_{a=1}^k \left(
X'_a\wedge R_a\right)
(dF,dG,dH)\ ,
\end{equation}
where $R_a$ is the expression (\ref{eq:neweq}).
The 3-form in the left-hand side, thanks to (\ref{3.5b}), is the exterior
derivative $d\omega''$ of the 2-form $\omega''$. Hence we have the identity
\begin{equation}
d\omega''(X_F,X_G,X_H) =\sum_{a=1}^k \left(X'_a\wedge R_a\right)
(dF,dG,dH)\ ,
\end{equation}
which proves the theorem.\
\rule{5pt}{5pt}

\end{document}